# Edge Chemistry Effects on the Structural, Electronic, and Electric Response Properties of Boron Nitride Quantum Dots

Dana Krepel, Lena Kalikhman-Razvozov, and Oded Hod

Department of Chemical Physics, School of Chemistry, The Raymond and Beverly Sackler Faculty of Exact Sciences, Tel Aviv University, Tel Aviv 69978, Israel

Abstract

The effects of edge hydrogenation and hydroxylation on the relative stability and electronic properties of hexagonal boron nitride quantum dots (*h*-BNQDs) are investigated. Zigzag edge hydroxylation is found to result in considerable energetic stabilization of *h*-BNQDs as well as a reduction of their electronic gap with respect to their hydrogenated counterparts. The application of an external in-plane electric field leads to a monotonous decrease of the gap. When compared to their edge-hydrogenated counterparts, significantly lower field intensities are required to achieve full gap closure of the zigzag edge hydroxylated *h*-BNQDs. These results indicate that edge chemistry may provide a viable route for the design of stable and robust electronic devices based on nanoscale hexagonal boron-nitride systems.

Since the isolation and characterization of graphene in 2004,[1] two-dimensional (2D) materials have become the subject of extensive research, exploring their unique physical properties.[2-9] Recently, several inorganic 2D materials have emerged as promising players in the field of nanoscience and nanotechnology.[10-13] While pure graphene is a semimetal, its inorganic counterparts show a variety of electronic properties ranging from semiconductors[11, 13] to insulators.[12] Of particular interest are the lower-dimensional derivatives of these materials, such as the quasi-one-dimensional nanoribbons that can be viewed as elongated stripes cut out of the 2D crystal and quasi-zero-dimensional quantum dots (QDs). Here, the structure, [14-17] dimension,[5-7, 18] and edge chemistry[19-21] of the systems, as well as external fields,[22, 23] may serve as control knobs for tuning their energetic stability and electronic and magnetic properties. This opens the way for many potential applications in a variety of fields including spintronic devices,[19, 23-25] gas sensors,[26-28] and nano-composites.[29]

Being the inorganic analogue of graphene, 2D hexagonal boron-nitride (*h*-BN) and its lower-dimensional derivatives are attracting increasing attention from the scientific community.[30-36] *h*-BN nanoribbons are predicted to possess a finite band-gap[37] and it was recently shown that zigzag boron-nitride nanoribbons (zBNNRs) may present a half-metallic character.[38-41] By further reducing their dimensionality to form hexagonal boron-nitride quantum dots (*h*-BNQDs), quantum-confinement and edge effects become dominant factors dictating their electronic and magnetic properties.[42-44]

In the present study, we use first-principle computational methods to study the effects of edge chemistry on the structural and electronic properties of *h*-BNQDs and their response toward the application of external electric fields. To this end, we consider three rectangular *h*-BNQDs with varying aspect ratios having both hydrogen edge passivation and zigzag edge hydroxylation (see Fig. 1). We find that the latter edge decoration scheme results in considerable energetic stabilization of *h*-BNQDs as well as a reduction of their electronic gap with respect to their hydrogenated counterparts.[39] Furthermore, upon the application of an external in-plane electric field a monotonous decrease of the gap is obtained for both chemical decoration schemes, where significantly lower field intensities are required to achieve full gap closure of the zigzag edge hydroxylated systems.

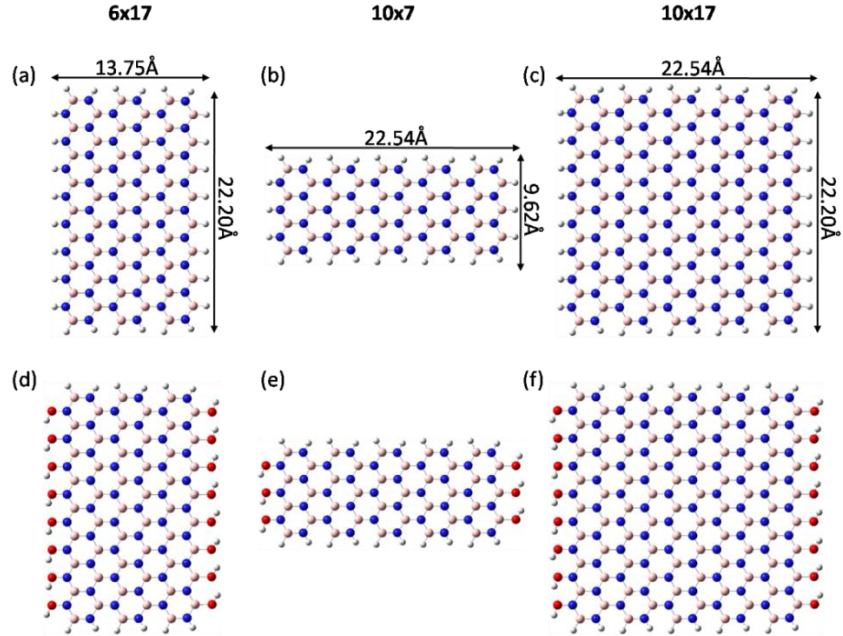

Figure 1. Optimized unit-cell geometries of the fully hydrogenated (upper panels) and zigzag edge hydroxylated (lower panels) 6x17 (a,d), 10x7 (b,e), and 10x17 (c,f) *h*-BNQDs studied. Here, the notation $N$x$M$ is used, where $N$ stands for the number of zigzag BN chains along the armchair edge and $M$ for the number of BN dimer chains along the zigzag edge.[7] Color code: blue, pink, gray, and red spheres represent nitrogen, boron, hydrogen, and oxygen atoms.

The calculations presented herein have been carried out using the GAUSSIAN suite of programs.[45] Geometry optimizations and electronic structure calculations have been performed within the B3LYP hybrid exchange-correlation density functional approximation[46] and the double-$\zeta$ polarized 6-31G$^{**}$ Gaussian basis set.[47] All structures have been fully relaxed in the absence of external fields. The relaxed coordinates have been centered on their respective centers of mass followed by a set of single-point calculations in the presence of such fields. The optimized coordinates of the various systems studied, at the B3LYP/6-31$^{**}$ level of theory, can be found in the supplementary information. Basis set convergence tests have been performed for the hydrogenated 6x17 *h*-BNQD (see Tab. 1) indicating that the relative structural stability energies and calculated electronic gaps are converged to within 3% and 2%, respectively, with respect to the choice of basis set.

| Basis set | 3-21G | 6-31G** | 6-311G** |
|---|---|---|---|
| δG (meV/atom) | 59 | 63 | 61 |
| Electronic gap (eV) | 5.78 | 5.96 | 5.92 |

Table 1. Basis set convergence test. Comparison of δG values, calculated using Eq. (1) below, and ground state Kohn-Sham electronic gaps of the hydrogenated 6x17 *h*-BNQD as calculated using the B3LYP exchange-correlation density functional approximation and the 3-21G, 6-31G$^{**}$ and 6-311G$^{**}$ Gaussian basis sets.

We start by studying the relative energetic stability of the different *h*-BNQDs. As these structures have different chemical compositions, the cohesive energy per atom does not provide a suitable measure for the comparison of their relative stability. Therefore, we adopt the approach used in Refs.[6, 39, 48-52] where one defines a zero-temperature Gibbs free energy of formation δ*G* for a *h*-BNQD as follows:

$$\delta G(\chi_H, \chi_O) = E(\chi_H, \chi_O) - \chi_H \mu_H - \chi_O \mu_O - \chi_{BN} \mu_{BN}. \tag{1}$$

Here, $E(\chi_H, \chi_O)$ is the cohesive energy per atom of a *h*-BNQD with given composition and dimensions, $\chi_i$ is the molar fraction of atom *i* (*i*=H, O, or BN pair) in the ribbon satisfying the relation $\sum_i \chi_i = 1$, and $\mu_i$ is the chemical potential of the $i^{th}$ constituent at a given state. We choose $\mu_H$ as the binding energy per atom of the singlet ground state of the hydrogen molecule, $\mu_O$ as the binding energy per atom of the triplet ground state of the oxygen molecule, and $\mu_{BN}$ as the cohesive energy per BN pair of a single 2D *h*-BN sheet, all calculated at the same level of theory as $E(\chi_H, \chi_O)$. This definition allows for an energy comparison between *h*-BNQDs with different compositions, where negative values represent stable structures with respect to the constituents. For all *h*-BNQDs considered, we compare the closed-shell singlet with the triplet spin state in order to identify the lowest energy spin configuration of each system. Unlike the case of graphene QDs (GQDs), we were not able to obtain an open shell singlet state for any of the systems studied.

In Fig. 2, we present the relative stabilities of the different *h*-BNQDs studied. Consistent with the results presented in Ref. [43], we find that, for all systems considered, the non-magnetic closed-shell singlet spin state is more energetically stable than the corresponding triplet state.

Generally, the stability of the quantum dots increases with size. For the hydrogen passivated $h$-BNQDs, $E_{10x17}^{\text{singlet}} - E_{6x17}^{\text{singlet}} = -12$ meV/atom and $E_{10x17}^{\text{singlet}} - E_{10x7}^{\text{singlet}} = -25$ meV/atom. For the zigzag edge hydroxylated systems these differences are $-9$ and $-94$ meV/atom, respectively. Similar to the case of zigzag graphene nanoribbons[49] and zBNNRs,[39] zigzag edge hydroxylation leads to a considerable energetic stabilization of the QD structure. This enhanced stability is attributed partially to the hydrogen bonds chain formed between adjacent edge hydroxyl groups (see Fig. 1(d)-(f)). Hence, the zigzag hydroxylated 6x17 and 10x17 systems are considerably more stable than the 10x7 $h$-BNQD, which has a shorter zigzag edge and thus forms shorter hydrogen bond chains.

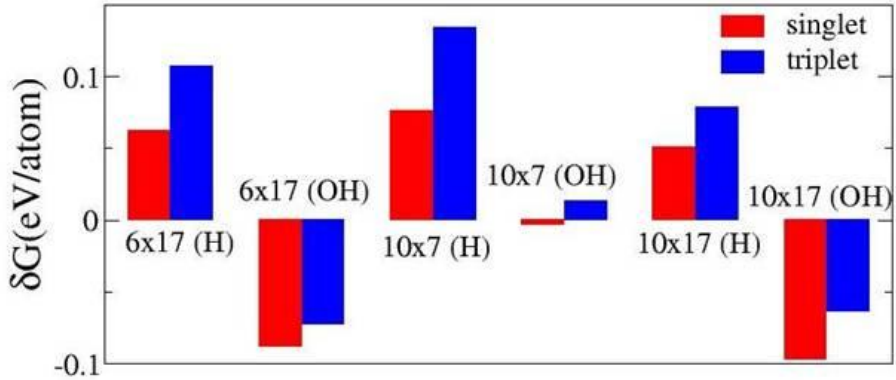

Figure 2: Ground state relative stabilities of the different edge hydrogenated and hydroxylated systems studied (see Fig. 1) obtained via Eq. (1) at the B3LYP/6-31G$^{**}$ level of theory. Negative values indicate stable structures with respect to the constituents. Results of the closed-shell singlet and triplet spin state calculations are presented in red and blue bars, respectively.

We now turn to study the electronic properties of the $h$-BNQDs considered. In Fig. 3, we present the energy gaps calculated as differences between the lowest unoccupied- (LUMO) and highest occupied- (HOMO) Kohn-Sham molecular orbital eigen-energies calculated at the B3LYP/6-31G$^{**}$ level of theory. All hydrogen passivated $h$-BNQDs present HOMO-LUMO gaps in the range of 5.60-5.96 eV. Upon hydroxylation, the gap drops by 20-30% for all systems studied suggesting that edge chemistry may prove as valuable tool for tailoring the electronic properties of boron-nitride based nanostructures.

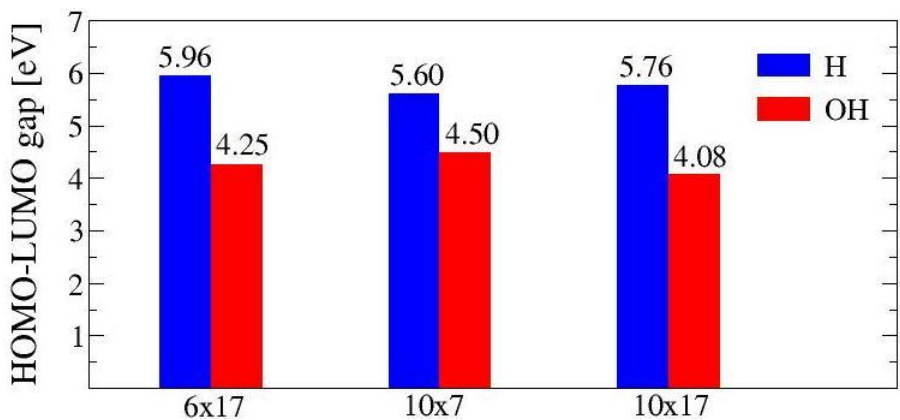

Figure 3: HOMO-LUMO Kohn-Sham gap response toward zigzag edge hydroxylation of the various *h*-BNQDs studied calculated at the B3LYP/6-31G$^{**}$ level of theory.

To rationalize this, we compare, in Fig. 4, the Kohn-Sham eigenvalues spectra of the hydrogenated and zigzag hydroxylated *h*-BNQDs. For all systems considered, it can be seen that hydroxylation leads to an up-shift of the occupied subspace, strongest for the HOMO orbital, and a relatively minor effect on the unoccupied orbital manifold resulting in the notable reduction of the HOMO-LUMO gap discussed above. When viewing the HOMO and LUMO orbitals (see right panels of Fig. 4 and middle panels of Fig. 8) it can be seen that the HOMO orbital is localized near the negatively charged nitrogen rich edge whereas the LUMO orbital localizes near the positively charged boron rich edge. Furthermore, upon careful examination it is evident that the contribution of the N-terminus hydroxyl groups to the HOMO orbital is larger than the corresponding contribution of the B-terminus hydroxyl groups to the LUMO orbital explaining why hydroxylation affects more the HOMO orbital energy. We attribute this to the fact that the matching between the HOMO orbital energy of the oxygen atom and that of nitrogen is much better than with boron as indicated by their ionization potentials (IP$_O$=13.6181 eV, IP$_N$=14.5341 eV, and IP$_B$=8.2980 eV),[53] thus leading to enhanced hybridization at the nitrogen rich edge. This is further reflected in the corresponding calculated N-O and B-O bond lengths of 1.36 Å and 1.54 Å, respectively, indicating that the hydroxyl group forms a stronger bond with the nitrogen rich edge than with its boron counterpart.

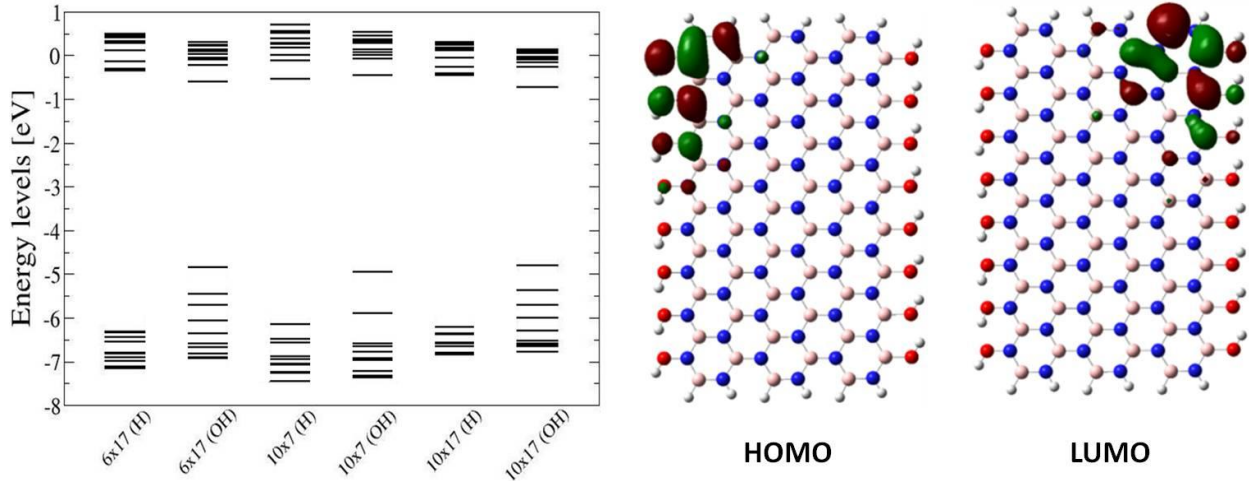

Figure 4: Left - Graphical representation of the low energy Kohn-Sham eigenvalues spectra of the various systems studied as calculated at the B3LYP/6-31G$^{**}$ level of theory. Right – HOMO and LUMO orbital isosurfaces of the 6x17 zigzag edge hydroxylated system. Isosurface value used is 0.02 electrons/Å$^3$.

From the analysis presented above we find that, while edge hydroxylation has a considerable stabilizing effect on *h*-BNQDs, its effect on their electronic properties is noticeable, yet somewhat limited. Despite the induced HOMO-LUMO gap reduction presented in Figs. 3 and 4 the hydroxylated *h*-BNQDs studied remain insulating with gaps exceeding 4 eV. Hence, we would like to identify complementary routes for controlling the electronic properties of these systems. To this end, we recall that for bare and edge-passivated zBNNRs a rich spectrum of electronic characteristics, ranging from metallic, through half-metallic, to semiconducting, is expected to occur under the influence of external electric fields.[38, 39] To examine whether *h*-BNQDs present a similar behavior, we apply electric fields of varying intensities along the zigzag and armchair axes of the hydrogenated and zigzag hydroxilated systems considered. Since all *h*-BNQDs studied lack mirror symmetry with respect to their central zigzag axis and the hydroxylated systems also lack mirror symmetry around the central armchair axis, we consider both positive and negative electric field polarities along each axis.

In Fig. , the HOMO-LUMO gap of the hydrogen passivated and zigzag hydroxylated *h*-BNQDs as a function of the intensity and polarity of an external electric field applied along the zigzag (upper) and armchair (lower) axes is presented. In all cases considered the application of the field results in a notable monotonous reduction of the gap regardless of its direction and polarity. These results clearly demonstrate that an external field can be used as a robust and efficient tool to control the electronic properties of *h*-BNQDs of nanoscale dimensions. Since the hydrogen passivated *h*-BNQDs do possess mirror symmetry around their central armchair axis a symmetric HOMO-LUMO gap decrease occurs when the field is applied along their zigzag axis (black diamond signs in the upper panels). Here, a very similar behavior is found for the 6x17 and 10x17 systems, which share the same zigzag edge length, where the gap practically vanishes at a field intensity of 1.7 V/Å (10x17) and 1.8 V/Å (6x17). For the 10x7 system, on the other hand, the sensitivity of the gap towards the external field is considerably weaker due to the shorter zigzag edge length, which limits the potential developed across the zigzag width of the dot to lower values.

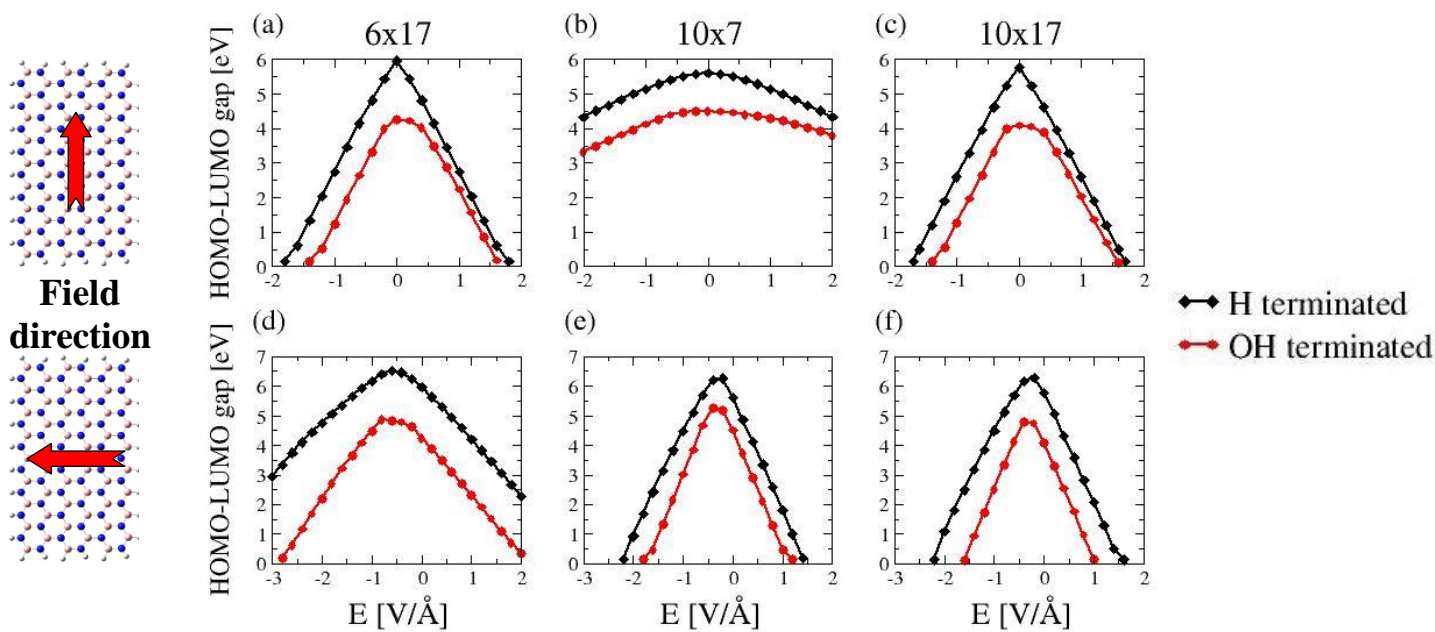

Figure 5: HOMO-LUMO gap as a function of the external electric field intensity of the 6x17 (left panels), 10x7 (middle panels) and 10x17 (right panels) hydrogenated (black diamond signs) and zigzag hydroxylated (red circles) *h*-BNQDs obtained at the B3LYP/6-31G$^{**}$ level of theory. The field is applied along both the zigzag (upper panels) and armchair (lower panels) axes as indicated by the images on the left.

When the field is applied along the armchair axis of the fully edge-hydrogenated systems (black diamond signs in the lower panels of Fig. ) the polarity symmetry is broken and the HOMO-LUMO gap peak is shifted towards the negative field region. This may be rationalized by the fact that the B- and N-edges of the *h*-BNQDs are oppositely charged and therefore an inherent electric field, pointing from the B to the N edge is induced in the system. The gap peak is thus obtained at the opposite external field intensity that nullifies the internal intrinsic filed. Since the partial charges along the B- and N-edges of all hydrogenated systems considered is quite similar, yielding Mulliken(Hirshfeld) charges on the B and N atoms of 0.36(0.14) and -0.54(-0.13), respectively (see upper panels of Fig. 6), it is mainly the length of the armchair edge that determines the intensity of the intrinsic electric field. Hence, the gap peak of the narrower 6x17 system appears at a larger external field intensity (-0.56 V/Å) than the wider 10x7 and 10x17 systems (-0.28 V/Å). In fact, we find that the ratio between these peak locations (2) corresponds well to the ratio between the armchair edge lengths of these systems (~1.6) indicating that the intrinsic field is indeed proportional to the armchair edge length and providing a quantitative understanding of this phenomena. Furthermore, the overall sensitivity towards the externally applied field is similar for the 10x7 and 10x17 systems that share the same armchair edge length and weaker for the 6x17 system that is narrower along this axis. As discussed above, this can be rationalized by the weaker potential developed across the narrower armchair width of the latter system at the same external field intensity.

When considering the zigzag hydroxylated systems no mirror symmetry exists for both the central armchair and zigzag axes and hence the polarity symmetry disappears in all cases (see red curves in Fig. ). Nevertheless, the general effect of the external field resembles that of the hydrogen passivated systems with overall smaller electronic gaps and lower field intensities required to achieve full gap closure. The fact that, when applying the field along the armchair axis, the peak gap shift is almost the same in the hydrogenated and hydroxylated systems indicates that the internal field due to the B-N edge polarization is similar in both systems. Indeed, when viewing the effective atomic charges (see Fig. 6) it can be seen that despite the individual effective atomic charge variations occurring upon hydroxylation the overall effective edge charge remains very similar to that of the hydrogenated system.

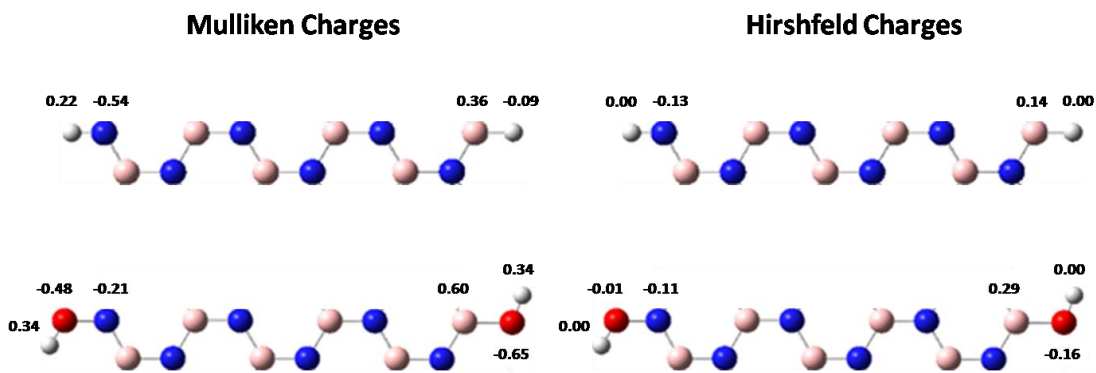

Figure 6: Effective edge atomic Mulliken (left panels) and Hirshfeld (right panels) charges of the central stripe of the hydrogenated (upper panels) and zigzag edge hydroxylated (lower panels) 6x17 system.

In order to gain better understanding of the field induced HOMO-LUMO gap variations described above, we plot, in Fig. 7, the field dependent Kohn-Sham HOMO and LUMO energies for the various systems studied. When the field is applied along the zigzag axis (upper panels) of the hydrogen passivated systems (black and brown diamond signs) the gap decrease results from a simultaneous decrease of the LUMO energy and increase of the HOMO energy up to the point where they meet. Both the HOMO and LUMO orbital energies of the 10x7 system present lower sensitivity towards the intensity of external field due to the smaller potential gradient that it induces, in this case, resulting in the overall weaker sensitivity presented in Fig. (b). A similar behavior is observed when the field is applied along the armchair edge of these systems (lower panels) with a shift of the minimum HOMO and maximum LUMO energies towards the negative external field values due to the intrinsic net field induced by the ribbon in this direction. As discussed above, here, the HOMO and LUMO orbital energies of the 6x17 system present weaker sensitivity towards the intensity of the external field.

Zigzag edge hydroxylation appears to have a stronger effect on the HOMO orbital energies than on the LUMO counterparts, as was shown in Fig. 4 for the field free case. When the field is applied along the zigzag axis (upper panels) the LUMO orbital energies (orange circles) behave very similar to those of the hydrogen passivated systems (brown diamond signs), whereas the HOMO orbital energies are considerably up-shifted (red circles vs. black diamond signs), especially for negative field intensities, thus reducing the overall gap values. A similar behavior is observed when the field is applied along the armchair axis (lower panels) with an additional decrease of LUMO orbital energies in the negative field polarity and an increase of the HOMO orbital energies in the positive polarity with respect to those of the hydrogenated systems.

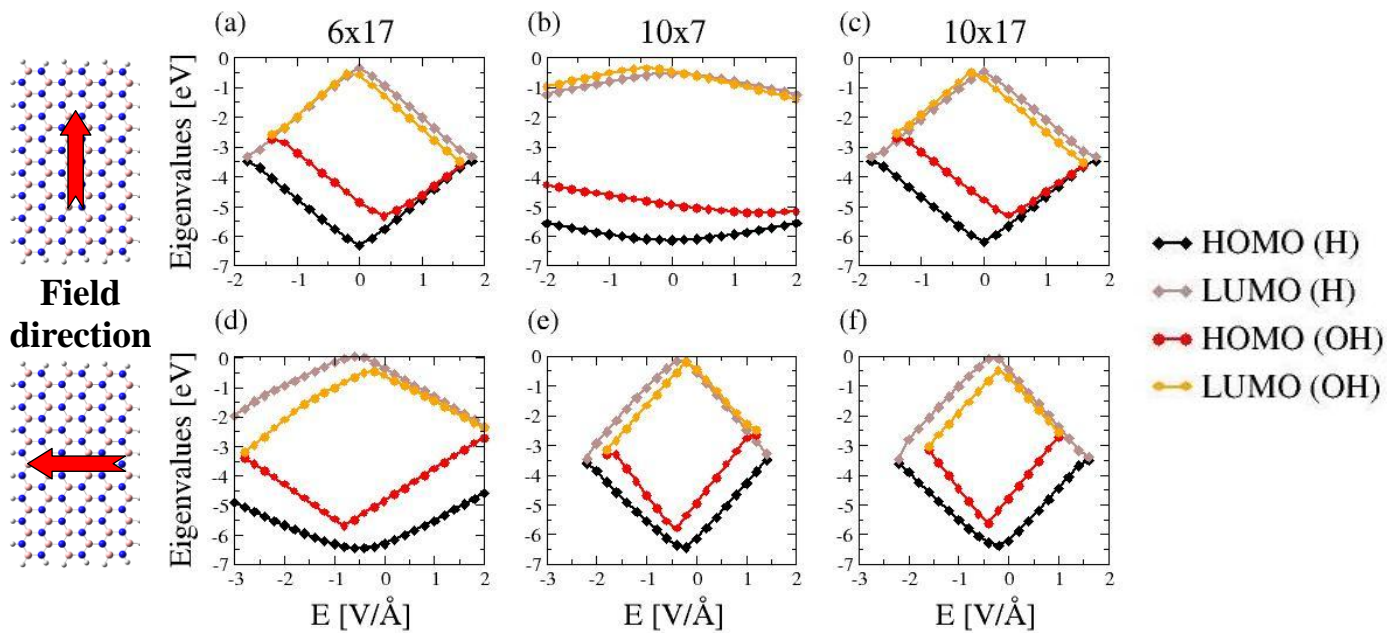

Figure 7: Eigenvalues of the Kohn-Sham HOMO and LUMO of the hydrogen passivated (black and brown diamonds, respectively) and zigzag hydroxylated (red and orange circles, respectively) 6x17 (left panels), 10x7 (middle panels), and 10x17 (right panels) *h*-BNQDs as a function of the intensity and polarity of the external electric field applied along their zigzag (upper panels) and armchair (lower panels) axes as obtained at the B3LYP/6-31G$^{**}$ level of theory.

To further explain these results we plot the HOMO and LUMO orbitals of the hydrogen passivated 6x17 system in the absence and presence of the external field applied along the zigzag (Fig. 8) and armchair (Fig. 9) axes. In the absence of an external field, the HOMO and LUMO orbitals symmetry with respect to a mirror operation along the central armchair axis and its asymmetry with respect to a similar operation along the central zigzag axis, are clearly reflected. The HOMO orbital spreads along the more electronegative nitrogen rich edge, whereas the LUMO orbital is distributed around the less electronegative boron rich edge. The application of an external field along the zigzag axis in the positive polarity strongly localizes both orbitals where the HOMO orbital localizes on the lower part of the nitrogen-rich edge (lower right panel of Fig. 8) and the LUMO orbital localizes around the upper section of the boron-rich edge (upper right panel of Fig. 8). We attribute the opposite behavior of these orbitals to the fact that the HOMO orbital resides on an electron rich edge and is repelled by the external field, whereas the LUMO orbital resides on a hole rich edge and thus follows the field. When the polarity of the field is reversed (left panels of Fig. 8) the resulting HOMO and LUMO orbitals are merely mirror reflections along the central armchair axis of those obtained for the positive field. The similar and pronounced behavior of the HOMO and LUMO orbitals under the influence of the external field thus explains why both orbital energies are strongly affected by the presence of the field as depicted in panel (a) of Fig. 7.

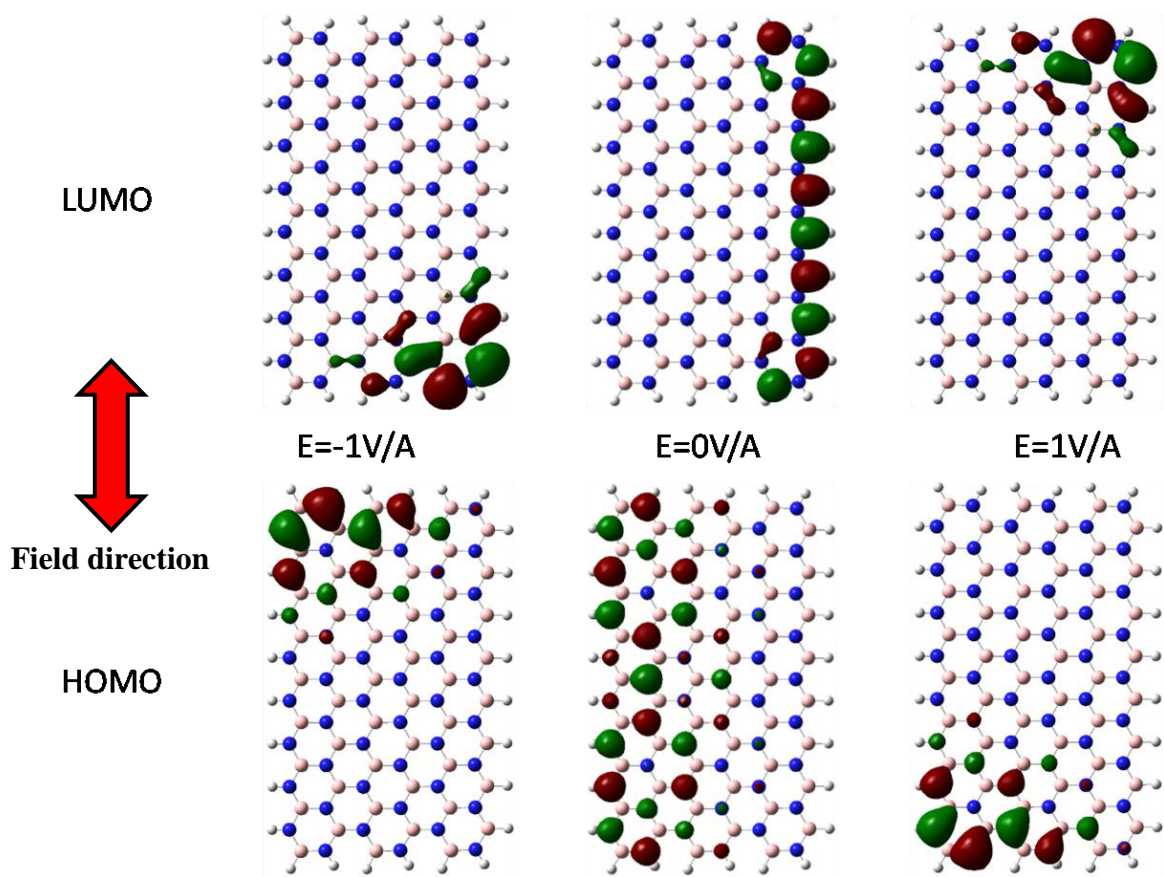

Figure 8: LUMO (upper panels) and HOMO (lower panels) orbital isosurfaces of the 6x17 *h*-BNQD as obtained at the B3LYP/6-31G$^{**}$ level of theory with an external electric field intensity of -1 V/Å (left panels), 0 V/Å (middle panels), and +1 V/Å (right panels) applied along the zigzag axis. The isovalue used in all panels is 0.02 electrons/Å$^3$.

When the field is applied along the armchair axis of the 6x17 system (Fig. 9) the axial nature of the orbitals is preserved and thus the resulting localization of both the HOMO and LUMO orbitals is much less pronounced. At the positive polarity, the localization of the HOMO (LUMO) orbital around the nitrogen (boron) rich edge merely slightly increases. At the reversed polarity the HOMO (LUMO) orbital migrates towards the boron (nitrogen) rich edge with similar localization as the field-free orbital at the field intensity presented. This further explains the weaker sensitivity of both HOMO and LUMO orbitals towards the external field that was presented for this system in panel (d) of Fig. 7. We note that a similar analysis for the hydrogen passivated 10x7 and 10x17 *h*-BNQDs produces analogous results.

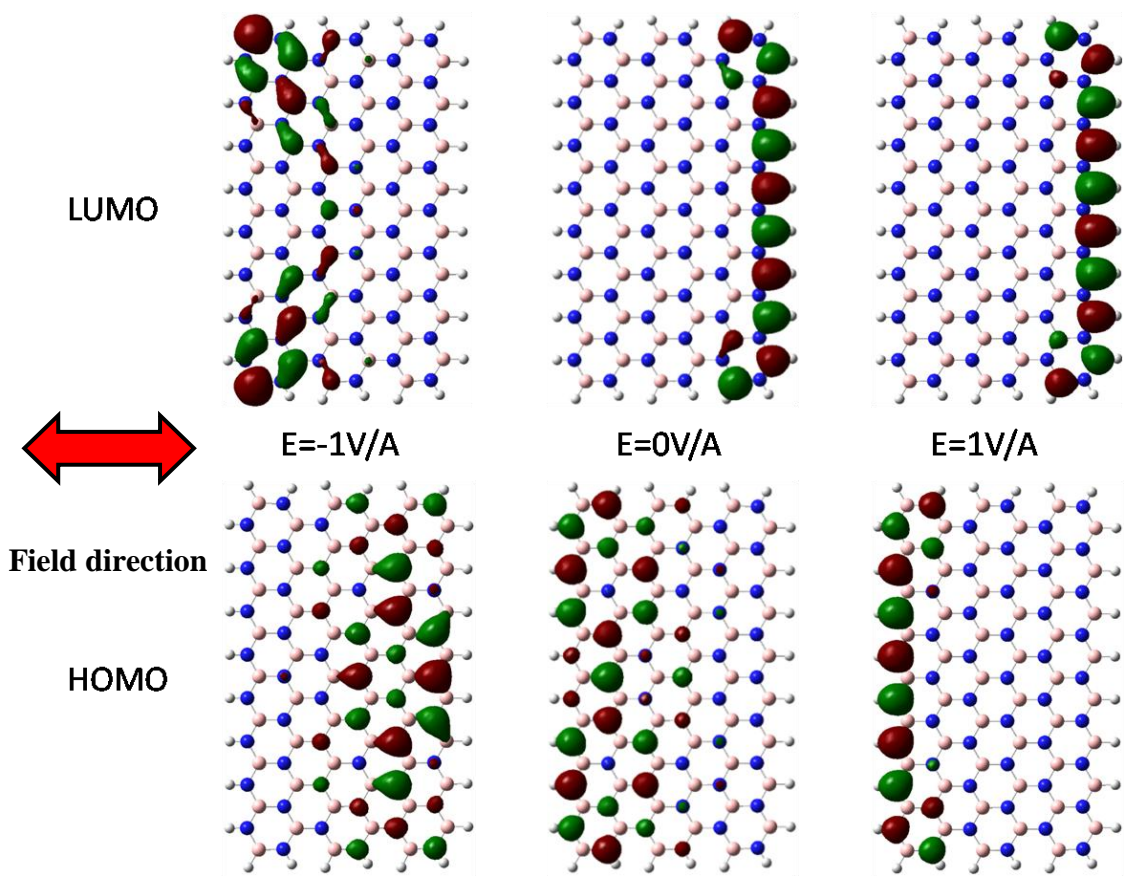

Figure 9: LUMO (upper panels) and HOMO (lower panels) orbitals of the 6x17 *h*-BNQD as obtained at the B3LYP/6-31G$^{**}$ level of theory with an external electric field intensity of -1 V/Å (left panels), 0 V/Å (middle panels), and +1 V/Å (right panels) applied along the armchair axis. The isovalue used in all panels is 0.02 electrons/Å$^3$.

In summary, we have studied the relative stability, electronic properties, and response to external electric field perturbations of *h*-BNQDs with both hydrogen and hydroxyl edge chemical decoration schemes. It was found that hydroxylation of the zigzag edges of the *h*-BNQD results in considerable energetic stabilization of the system. The application of external in-plane electric fields was found to alter the electronic properties of the system resulting in a pronounced reduction of the HOMO-LUMO gap up to full gap closure depending on the *h*-BNQD's dimensions, chemical edge decoration, and orientation of the external field with respect to the main symmetry axes of the system. This opens the way for the design of robust nanoelectronic components, presenting novel functionalities, based on edge functionalized *h*-BNQDs.

Acknowledgments

This work was supported by the Israel Science Foundation (ISF) under Grant No. 1313/08, the Center for Nanoscience and Nanotechnology at Tel Aviv University, the Israeli Ministry of Defense, and the Lise Meitner-Minerva Center for Computational Quantum Chemistry. D. K. acknowledges the Miriam and Aaron Gutwirth Excellence Scholarship Award.